 \documentclass[preprint2]{aastex}
 \usepackage{amsmath}

\newcommand{\teff}{$T_{\rm eff}$}
\newcommand{\logg}{log$g$}
\newcommand{\feh}{[Fe/H]}
\newcommand{\afe}{[$\alpha$/Fe]}
\newcommand{\R}{$R$}

\newcommand{\MK}{$M_K$}
\newcommand{\rg}{$R_g$}
\newcommand{\vrad}{$v_R$}
\newcommand{\vphi}{$v_\phi$}
\newcommand{\vz}{$v_z$}
\newcommand{\kms}{km s$^{-1}$}
\newcommand{\zmax}{$z_{max}$}

\newcommand{\lzlc}{$L_z/L_c$}
\newcommand{\ecc}{$ecc$}
\newcommand{\lz}{$L_z$}

\newcommand{\ns}{\emph{narrow stripe}}
\newcommand{\mn}{\emph{main sample}}

\slugcomment{}

\shorttitle{The K giant stars from the LAMOST survey data II}
\shortauthors{Liu et al.}

\begin{document}


\title{The K giant stars from the LAMOST survey data II: the Hercules stream in radial migration}


\author{Chao Liu\altaffilmark{1}, Xiao-Yan Chen\altaffilmark{1}, Jun Yin\altaffilmark{2}, Bo Zhang\altaffilmark{1}, Li-Cai Deng\altaffilmark{1}, Yong-Hui Hou\altaffilmark{3}, Zheng-Yi Shao\altaffilmark{2}, Jun-Chen Wan\altaffilmark{1}, Hai-Feng Wang\altaffilmark{1}, Yue Wu\altaffilmark{1}, Yu Xin\altaffilmark{1}, Yan Xu\altaffilmark{1}, Ming Yang\altaffilmark{1}, Yong Zhang\altaffilmark{3}}
\email{liuchao@nao.cas.cn}

\altaffiltext{1}{Key Laboratory of Optical Astronomy, National Astronomical Observatories, Chinese Academy of Sciences, Datun Road 20A, Beijing 100012, China}
\altaffiltext{2}{Shanghai Astronomical Observatory, Chinese Academy of Sciences, Shanghai 200030, China}
\altaffiltext{3}{Nanjing Institute of Astronomical Optics \& Technology, National Astronomical Observatories, Chinese Academy of Sciences, Nanjing 210042, China}


\begin{abstract}
We estimate the age for the individual stars located at the lower part of the red giant branch from the LAMOST DR2 K giant sample. Taking into account the selection effects and the volume completeness, the age--metallicity map for the stars located between 0.3 and 1.5\,kpc from the Sun is obtained. A significant substructure (denoted as the \ns) located from (age, \feh)$\sim$(5, 0.4) to (10\,Gyr, -0.4\,dex) in the age--metallicity map is clearly identified. Moreover, the \ns\ stars are found the dominate contributors to several velocity substructures, including the well-known Hercules stream. The substantially large difference between the observed guiding-center radii and the birth radii inferred from the age--metallicity relation is evident that the \ns\ stars have been radially migrated from about \R$\sim4$\,kpc to the solar neighborhood. This implies that the Hercules stream may not be owe to the resonance associated with the bar, but may be the kinematic imprint of the inner disk and later moved out due to radial migration.
We estimate that the traveling speed of the radial migration are roughly 1.1$\pm0.1$ kpc Gyr$^{-1}$, equivalent with about $1.1\pm0.1$\,\kms. This is in agreement with the median \vrad\ of $2.6^{+1.8}_{-1.9}$\,\kms of the \ns. We also obtain that about one third stars in the solar neighborhood are radially migrated from around 4\,kpc. Finally, we find that the radial migration does not lead to additional disk thickening according to the distribution of \zmax.
\end{abstract}





\section{Introduction}\label{sect:intro}
Although the non-axisymmetric structures, e.g. spiral arms or bar, in the Galactic disk only occupy a small fraction of the total disk mass, they may drive significant re-distributions of the angular momentum, energy, and mass of the stellar disk in the secular evolution. \citet{sellwood02} have demonstrated that the transient spiral structures can excite strong exchange of the angular momenta between the stars at around the corotation radius. When the spiral structures disappear, the stars gaining (losing) angular momenta will radially migrate outward (inward).
Not only the transient spiral structures, the resonance overlap of the spiral arms and the central bar~\citep{minchev10} and the merging dwarf galaxies~\citep{bird12} are also able to drive the radial migration.

An immediate result in the solar neighborhood of the radial migration is a flat and broad age--metallicity relation~\citep[AMR,][]{sellwood02,roskar08,schoenrich09a,loebman11}, which has been already noticed from the observations~\citep{edvardsson93,bergemann14} for more than two decades. 
N-body simulations have shown that the radial migration can lead to other effects in chemodynamics of the disk. \citet{roskar08} found that not only the AMR is flat, the radial metallicity gradient also becomes flat for the old stars. \citet{loebman11} noticed that, in the old stellar populations, the radial migration can smear out the azimuthal velocity--metallicity anti-correlation, which is prominent in the young stellar populations due to the steeply radial metallicity gradient and the epicyclic motions.
In recent years, observations gradually unveil evidences for radial migration in the solar neighborhood. \citet{yu12} demonstrated that the radial metallicity gradient of the main-sequence stars is indeed steeper for young stars and flat for old stars. \citet{lee11} and~\citet{liu12} revealed that the azimuthal velocity--metallicity anti-correlation does disappear in the $\alpha$-enhanced stars, which should be quite old. 
Recently,~\citet{kordopatis15} found that the supersolar metallicity stars in the solar neighborhood is the latest evidence for radial migration. 
However, unlike the previous works, \citet{haywood13} claimed that no clear evidence of radial migration is found in their high resolution spectroscopic samples.

Radial migration is also associated with the origin of the Galactic thick disk. \citet{schoenrich09a,schoenrich09b} introduced a novel Galactic evolution model taking into account the radial migration and inferred that the thick disk is likely originated from radial migration. \citet{bovy12a} found from SEGUE~\citep{yanny09} G-dwarf stars that the thick disk should be continuously heated by internal mechanism, like radial migration, from the thin disk. \citet{liu12} also agreed that the radial migration may be at least partly responsible for the thick disk. However,~\citet{minchev12} found from their simulations that the radial migration does little with the thickening of the disk. As a response, \citet{roskar13} demonstrated that both heating and radial migration contribute to the disk thickening. And~\citet{solway12} showed that the migrating stars increase their vertical excursions and angular momenta simultaneously. This debate needs to be clarified with more observational evidence.

In order to well address the observational effect of the radial migration, particularly to estimate its efficiency in the disk evolution, several observational quantities are crucially required. First, the age of stars is always the most important quantity. Theoretical works has proved that the AMR is a powerful tool to address the Galactic evolution~\citep{sellwood02,roskar08,schoenrich09a,loebman11}. However, in practice, the age for individual stars is always the most difficult one to be determined. For high resolution spectroscopic data, the age has been determined only for a small sample with around 1000 stars~\citep{edvardsson93,haywood13,bensby14}, which is far from sufficient to unveil the details of the disk evolution. On the other hand, although large surveys have provided millions of low-resolution stellar spectra, few of them has the age been determined. Second, elemental abundances, e.g. \feh, \afe\ etc., are also very important for tracing the elemental enrichment during the evolution of a stellar population~\citep{lee11,liu12,bovy12a,bovy12b,bovy12c}. Third, the distribution function of stars, combined with age, can also reflect the dynamical evolution. As the projection of the distribution function, the three dimensional velocity distribution in the solar neighborhood has been linked with age~\citep{dehnen98b,quillen01,robin03,soubiran03,tian15}.
 
Since the values of the age and chemical elemental abundances have been well discussed in lots of other works, here we stress the importance of the velocity distribution.
It is noted that, in the solar neighborhood, the velocity distribution exhibits some puzzling substructures. Among them, the Hercules stream is the most prominent one~\citep{dehnen98,antoja12,antoja14,antoja15,xia15}. Abundance and age analysis disfavor that the Hercules stream is disrupted from supercluster or satellite galaxy~\citep{famaey05,bensby14}. \citet{dehnen00} compared this structure with the test particle simulation and found that it is quite similar with the resonant structure induced by the bar. \citet{antoja14} further extended the test from the distance of a few hundreds parsec to 1-2 kpc away from the Sun. Although~\citet{fux01} agreed that the Hercules stream is associated with the bar, he claimed that it is composed of the stars on hot and chaotic orbits from the bar rather than the nearby stars with resonant orbits. 
However, whether the substructures, such as the Hercules stream, are associated with the secular evolution has never been carefully discussed. 
 
In this work, we make use of the large amount of data from the LAMOST survey~\citep{cui12,zhao12,deng12}. We combine the AMR with the velocity distribution for about 17,000 low red giant branch (RGB) stars located within 1.5\,kpc around the Sun from the LAMOST DR2 K giant sample~\citep{liu14}. We demonstrate that even with low-resolution spectra, the age estimates for the low RGB stars are sufficiently good for the statistical study of the chemodynamical evolution of the Galactic disk. 
We unveil that the Hercules stream is part of the relatively old population radially migrated from the inner disk to the solar neighborhood. We also constrain the efficiency of the radial migration from the observational data.

The paper is organized as below. In section 2, we describe how the low RGB stars are selected. In section 3, we estimate the age for the selected RGB stars and assess its performance. The substructures, including the supersolar metallicity stars and the Hercules stream, are unveiled from the AMR and velocity distribution in section 4. In section 5, we infer that these substructures are radially migrated from the inner disk and provide estimates of the efficiency of the radial migration. Finally, the brief conclusions are drawn in section 6.

\section{Data}\label{sect:data}
\begin{figure}[htbp]
\centering
\includegraphics[scale=0.6]{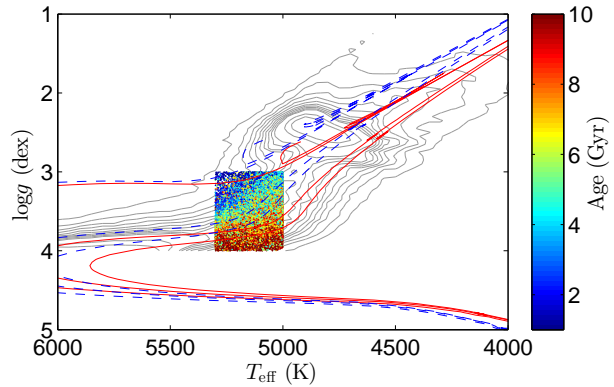}
\caption{The \teff\ vs. \logg\ distribution of the RGB sample (color dots) selected from the LAMOST DR2 K giant sample. The color codes the derived age. The gray contours indicate the complete K giant sample. The red solid lines are theoretical isochrones at 1, 5, and 9\,Gyr (from top to bottom) and \feh=0\,dex from PARSEC database \citep{parsec}. The blue dashed lines are the isochrones at the same ages but with \feh=-0.3\,dex. }\label{fig:sample}
\end{figure}

Not all stars are sensitive to age in the stellar parameter space. It is well known that the turn-off (TO) and sub-giant branch (SGB) stars are good tracers for age. However, the contaminations from the main-sequence stars to the TO stars, the complicated situation in stellar physics (evolution of binary systems, fast rotation, overshooting etc.) may give unclear systematics for the age estimation for TO stars~\citep[as an instance, see][]{li14}. And because the time scale is very short in SGB stage, not many stars are sampled in this region. Therefore, we do not select either of the two kinds of stars as the age tracer, but turn to use the RGB stars. The stars located at the top of the RGB are not sensitive to age and may be contaminated by the asymptotic giant branch stars. However, the low RGB stars, which located below the red clump and red bump stars, are perfect tracers for age. First, the stellar parameters of these RGB stars are very sensitive to the age, as shown in Figure~\ref{fig:sample}. The isochrones show that \logg\ can move by about 0.6\,dex when the age changes from 1 to 5\,Gyr for stars with \feh$=-0.3$ (blue dashed lines) and 0\,dex (red solid lines). Then, with the accuracy of $\sim0.2$\,dex in \logg, one can expect that, ideally, the uncertainty of the age estimate may be only around 2\,Gyr. Second, this region is very clean, contaminated by neither the red clump stars nor the main-sequence stars. Finally, the stars can stay in this stage for a few 100\,Myr, allowing more observed samples than SGB stars~\citep{BM98}. 

We adopt the LAMOST stellar parameters~\citep{wu11a,wu11b,wu14,luo15}, which have uncertainty of $\sim110$\,K, $\sim0.11$\,dex and $\sim0.19$\,dex for effective temperature, metallicity, and surface gravity, respectively~\citep{gao15}, and select the stars with  5000$<$\teff$<5300$\,K and $3<$\logg$<4$\,dex.
Note that~\citet{liu15} provided a more accurate \logg\ estimates ($\sim0.1$\,dex) based on the asteroseismology. However, the seismic scaling relation may not be validate for low metallicity stars, such as the thick disk stars. Therefore, we decide to adopt the LAMOST \logg\ estimates with slightly larger error but less limit of use. We select the stars with signal-to-noise ratio larger than 20 in their spectra to ensure that the accuracies of the stellar parameters are sufficiently good.
We also select the stars with \feh$>-1.2$\,dex to exclude most of the halo stars. From the LAMOST DR2 K giant sample, we find a total of 42166 RGB stars after applying all above criteria. Figure~\ref{fig:sample} shows the location of the selected RGB samples with color coded age estimates in \logg\ vs. \teff\ plane. 

\section{Age determination}\label{sect:agdetermin}
\begin{figure}[htbp]
\centering
\includegraphics[scale=0.5]{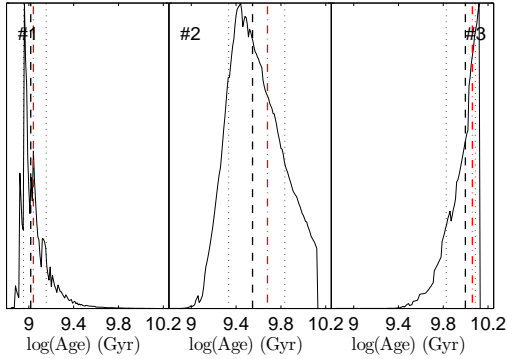}
\caption{The likelihood of logarithmic age for 3 test samples. Each panel corresponds to one mock star. The solid lines are the likelihoods of the log age, the black dashed lines and the dotted lines are the median and $\pm1\sigma$ values derived from the likelihoods, respectively, and the red dashed lines indicate the true log ages.}\label{fig:agesample}
\end{figure}

\subsection{Method}\label{sect:method}

\begin{figure}[htbp]
\centering
\begin{minipage}{9cm}
\centering
\includegraphics[scale=0.6]{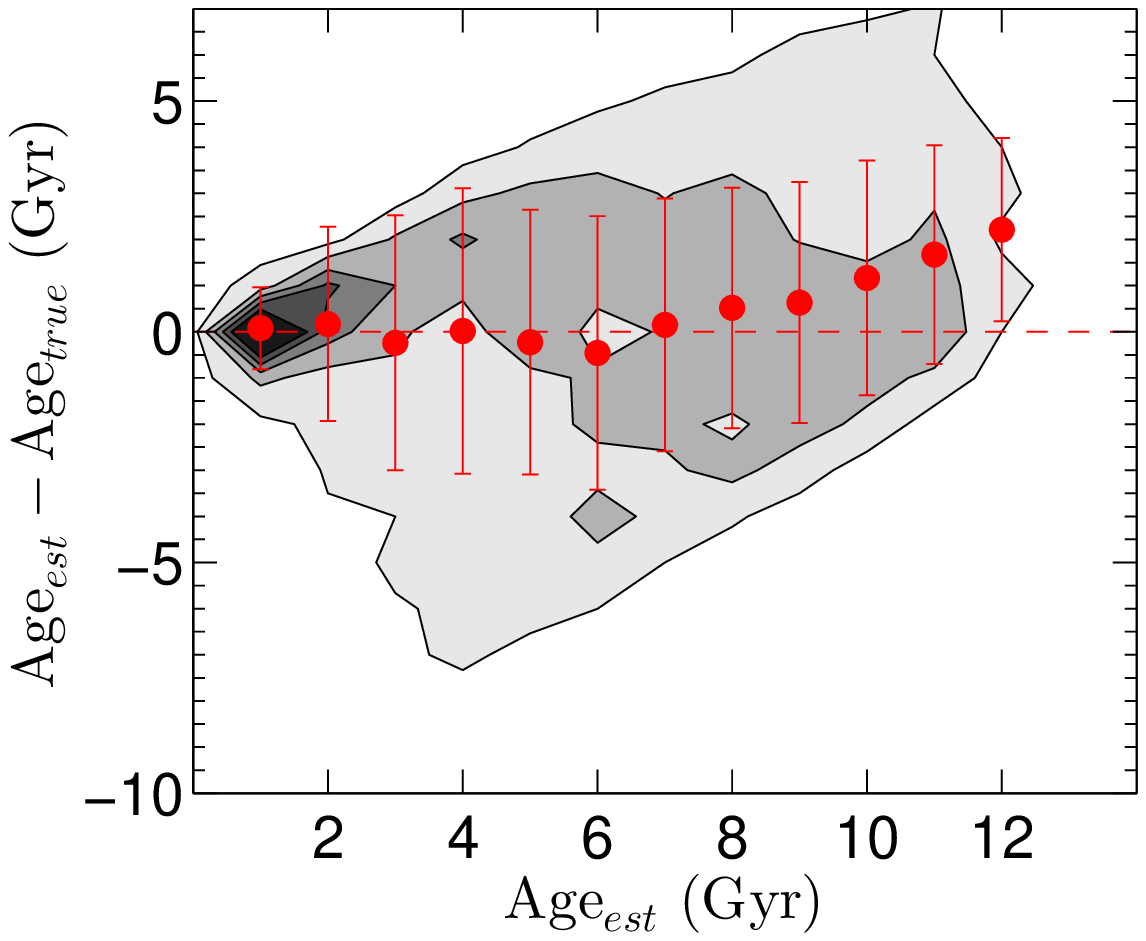}
\includegraphics[scale=0.6]{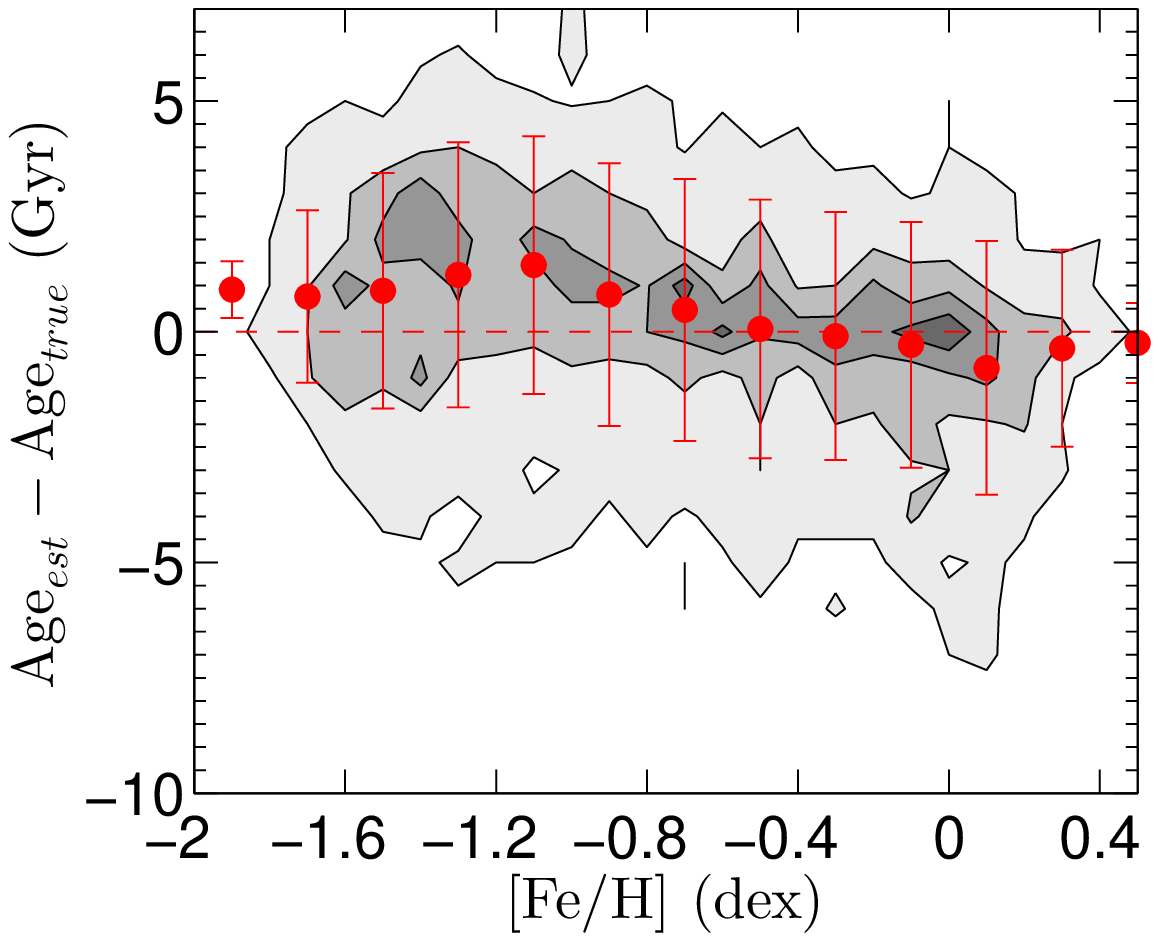}
\end{minipage}
\caption{The top panel shows the distribution of the residual of age vs. the estimated age for the test sample (contours). The red filled circles with error bars indicate the medians and dispersions (standard deviation) of the residuals at various age bins. The bottom panel shows the residual of age as a function of \feh\ for the test data. }\label{fig:ageerror}
\end{figure}

We select a set of theoretical isochrones from PARSEC database \citep{parsec} with fine grids in both age ($\Delta$loga=0.01) and \feh\ ($\Delta$[Fe/H]=0.1) to derive the age for the RGB sample. The synthetic data is clipped with $4950\leq$\teff$\leq5350$\,K and $2.9\leq$\logg$\leq4.1$ to well overlap the observed data. The age of a RGB star is estimated from the likelihood technique. The likelihood of the $i$th star to have the observed stellar parameters $O_{i,k}$ (in which $k=1,2,3$ represent for the observed \teff, \logg, and \feh, respectively) given the age ($\tau$), initial stellar mass ($\mathcal{M}_{ini}$), and absolute magnitude ($M$) is defined as
\begin{equation}\label{eq:likelihoodage}
\begin{aligned}
&L_{i}(O_i|\tau,\mathcal{M}_{ini},M)=exp\large(\\
&-\sum_{k=1}^{3}{(O_{i,k}-T_{k}(\tau,\mathcal{M}_{ini}, M))^2\over{2\sigma_{i,k}^2}}\large),
\end{aligned}
\end{equation}
where $\sigma_{i,k}$ is the uncertainties of the parameters, $T_{k}(\tau,\mathcal{M}_{ini}, M)$ stands for the \teff, \logg, and \feh\ of the given $\tau$, $\mathcal{M}_{ini}$, and $M$. The likelihood of the age is then determined by marginalizing the likelihood $L_{i}$ over absolute magnitude and initial stellar mass. Figure~\ref{fig:agesample} shows the likelihoods of the log age for 3 test samples generated in next subsection. The initial stellar mass for the selected RGB stars tightly depend on the age, because these stars stay only for a few 100\,Myr in this evolutionary stage. Therefore, unlike~\citet{serenelli13} and~\citet{bergemann14}, who took into account the initial mass function as a prior in their Bayesian technique, the prior of initial stellar mass helps little to constrain the age for our sample and thus is not included in equation~(\ref{eq:likelihoodage}).

\subsection{Performance of the age estimates}\label{sect:ageestperform}
We randomly draw about 3,000 test points within $5000<$\teff$<5300$\,K and $3<$\logg$<4$\,dex (allowing multiple sampling for a same point) from the isochrones covering the age up to 13\,Gyr. Given that the uncertainties of the stellar parameters are $\Delta_{{\rm log}g}=0.2$\,dex, $\Delta_{\rm [Fe/H]}=0.1$\,dex, and $\Delta_{T_{\rm eff}}=100$\,K and adopting that the random errors follow Gaussian distributions, we add arbitrary errors onto the test  data and hence create a mock catalog. Then, we estimate the age for this sample based on equation~(\ref{eq:likelihoodage}). Figure~\ref{fig:agesample} shows the likelihoods of the age for 3 test examples. 

Figure~\ref{fig:ageerror} shows the performance of the age for all test stars. The top panel displays that, statistically, the age estimates have the uncertainties of about 2\,Gyr, although a few old stars are underestimated by about 5\,Gyr and a few intermediate-age ones are overestimated by about 5\,Gyr. Moreover, averagely, the age estimates for the stars with age$_{est}>8$\,Gyr are slightly overestimated by at most 2\,Gyr. These systematics are naturally explained by the lower sensitivity to age in \logg\ vs. \teff\ plane for old stars. Indeed, in figure~\ref{fig:sample}, the isochrones at old ages are very concentrated, hence the age estimates for old stars are less accurate than the younger stars given the same uncertainties in \logg\ and \teff. 
 The bottom panel of figure~\ref{fig:ageerror} shows the correlation of the residual of age with \feh. It shows that the age may be overestimated by about 1-2\,Gyr, which is still within 1$\sigma$ of uncertainty, for the metal-poor stars with \feh$<-0.8$\,dex. 

\begin{figure}[htbp]
\centering
\begin{minipage}{9cm}
\centering
\includegraphics[scale=0.5]{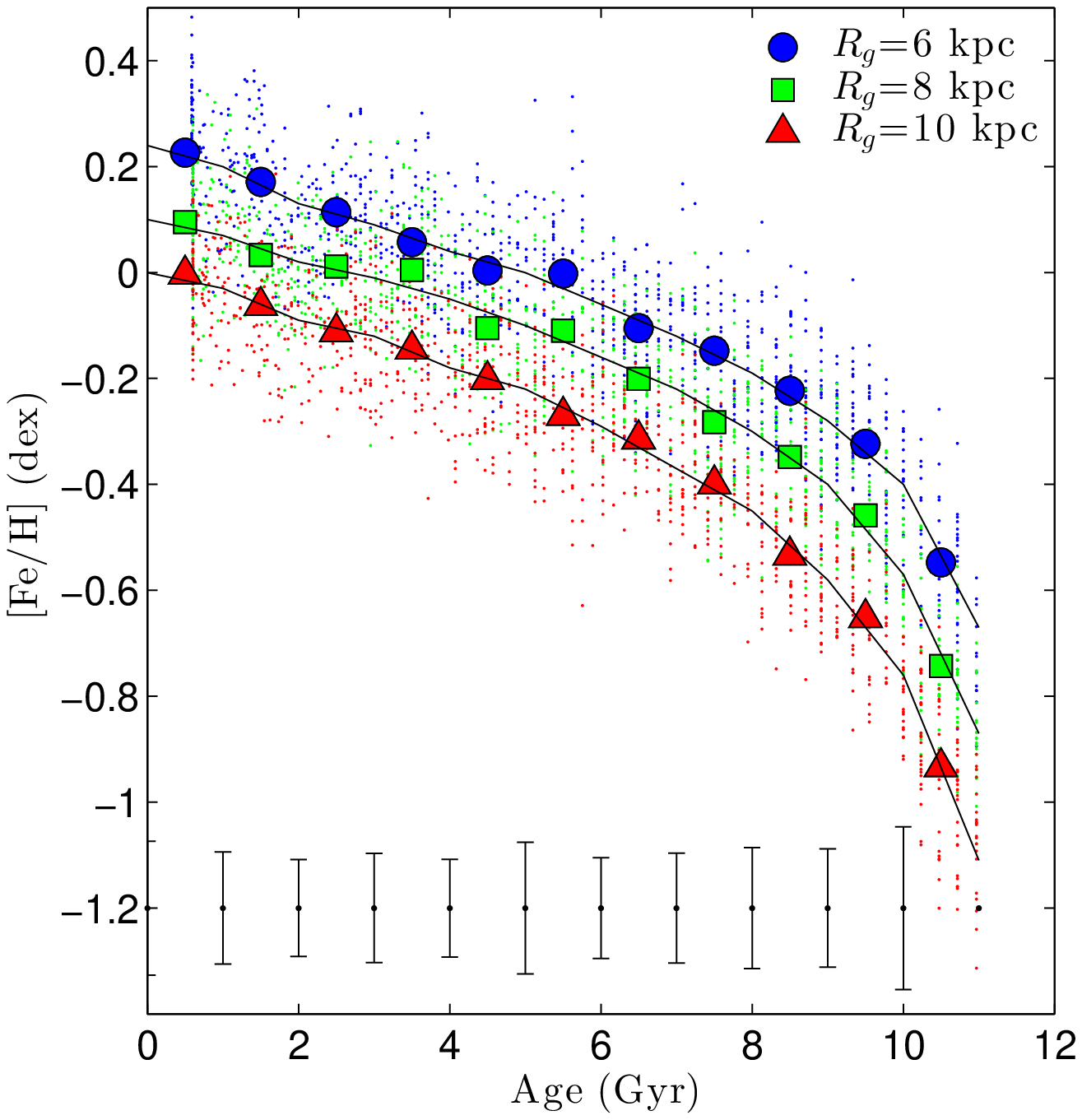}
\includegraphics[scale=0.5]{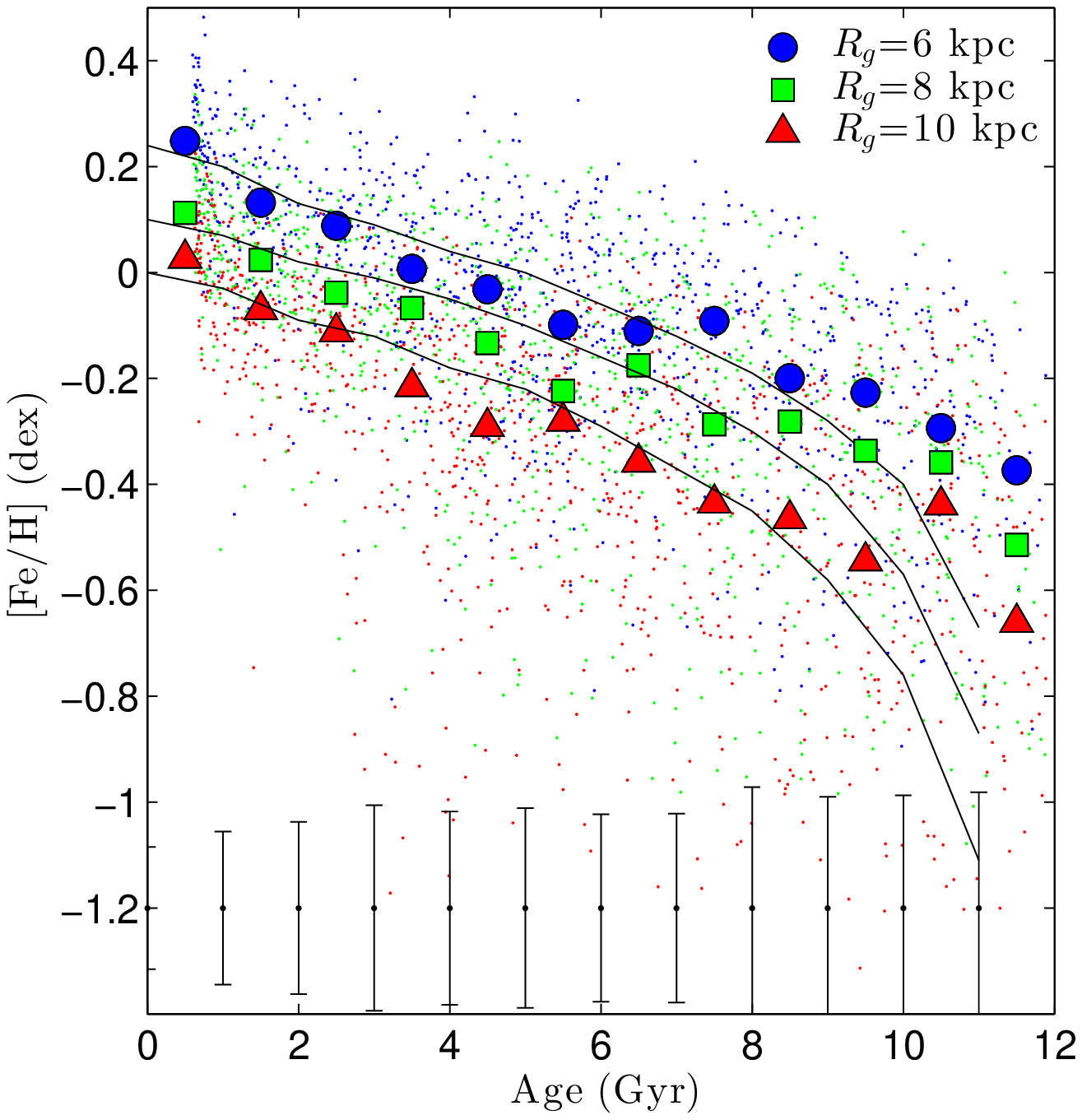}
\end{minipage}
\caption{In the top panel, the blue, green, and red dots represent for the mock stars with noise-added \feh\ and true ages located at \R$=6$, $8$, and $10$\,kpc, respectively. The large blue circles, green rectangles, and the red triangles are the mean values at each age bin for the three radii, respectively. The three black solid lines are the theoretical AMR from~\citet{minchev13} at \R$=6$, 8, and 10\,kpc from top to bottom, respectively. The mean dispersion of \feh\ at each age bin is shown at the bottom of the panel. The bottom panel shows the AMR for the same mock data but with the estimated age.}\label{fig:testAMR}
\end{figure}

In order to investigate how these systematics distort the AMR, we run another test. We create the test mock data from the theoretical AMR taken from~\citet[][]{minchev13} at Galactocentric radii of \R$=$6, 8, and 10\,kpc . We firstly arbitrarily select a point from the theoretical AMR.  Then we randomly select a pair of \teff\ and \logg\ within $5000<$\teff$<5300$ and $3<$\logg$<4$ from the isochrone with most likely age and metallicity in PARSEC. Adding to random errors in \teff, \logg, and \feh\ with uncertainties of 100\,K, 0.2\,dex, and 0.1\,dex, respectively, we generate 1000 mock data for each \R. The top panel of figure~\ref{fig:testAMR} shows the AMR for the 3000 mock data with noise-added \feh\ and the true age values. Without uncertainty of age, the mean \feh\ at each age bin (shown in blue filled circles, green rectangles, and red triangles for \rg$=6$, 8, 10\,kpc, respectively) follows the theoretical AMR (the black solid lines) quite well.  We then reconstruct the AMR with the estimated ages, as shown in the bottom panel of figure~\ref{fig:testAMR}. It is seen that the mean AMRs from the derived age are well in agreed with the real ones until $\sim8$\,Gyr. At larger age, due to the systematic bias appeared in figure~\ref{fig:ageerror}, the mean AMRs of the test data are more flatter than their true values. Therefore, one must be very careful when draw any conclusion with the stars in such old age. Although the AMR biases to higher metallicity at the old age regime, the anti-correlation between \rg\ and \feh, which is critically important for the investigation of radial migration, is still held. 

Another effect in the age estimates is that the dispersion of \feh\ at each age bin in the bottom panel increases by a factor of 2 than those in the top panel, implying that the uncertainty of the age can add extra 0.1\,dex dispersion in \feh\ in the age--metallicity (AM) map.

\subsection{The orbital properties}
In oder to derive the orbital properties of the stars, we cross-match these RGB stars with UCAC4~\citep{ucac4} and PPMXL~\citep{ppmxl} catalogs to obtain the proper motions. With the distance derived by~\citet{carlin15}, we are able to estimate the 3-D velocities in the cylindrical coordinates. Then, we adopt the same Galactic gravitation potential as~\citet{liu12} and apply the same techniques to derive the guiding center radius, \rg, the maximum orbital height, \zmax, the eccentricity, \ecc, and the circularity, \lzlc\footnote{\rg\ is derived from the angular momentum, \lz. \zmax\ is obtained from the orbital integration. \ecc\ is defined as $ecc=(R_{apo}-R_{peri})/(R_{apo}+R_{peri})$, where $R_{apo}$ and $R_{peri}$ are the apo- and peri-center of the stellar orbit. $L_c$ is the maximum angular momentum at the location of the star given that the orbit is exactly circular.} based on the proper motions from UCAC4 and PPMXL, respectively. In order to select the stars with relatively reliable proper motions, we remove the stars with the following criteria:
\begin{enumerate}
\item Difference in \rg\ derived from UCAC4 and PPMXL is larger than 3$\sigma$.
\item Difference in \lzlc\ derived from UCAC4 and PPMXL is larger than 3$\sigma$.
\item The error of the proper motion in UCAC4 is larger than 5\,mas yr$^{-1}$.
\end{enumerate}
After the exclusion, 21961 stars are left in the sample. In the rest of this paper, we only use the orbital properties derived from the UCAC4 proper motions.

\subsection{Selection effect correction}
\begin{figure}[htbp]
\centering
\begin{minipage}{9cm}
\includegraphics[scale=0.55]{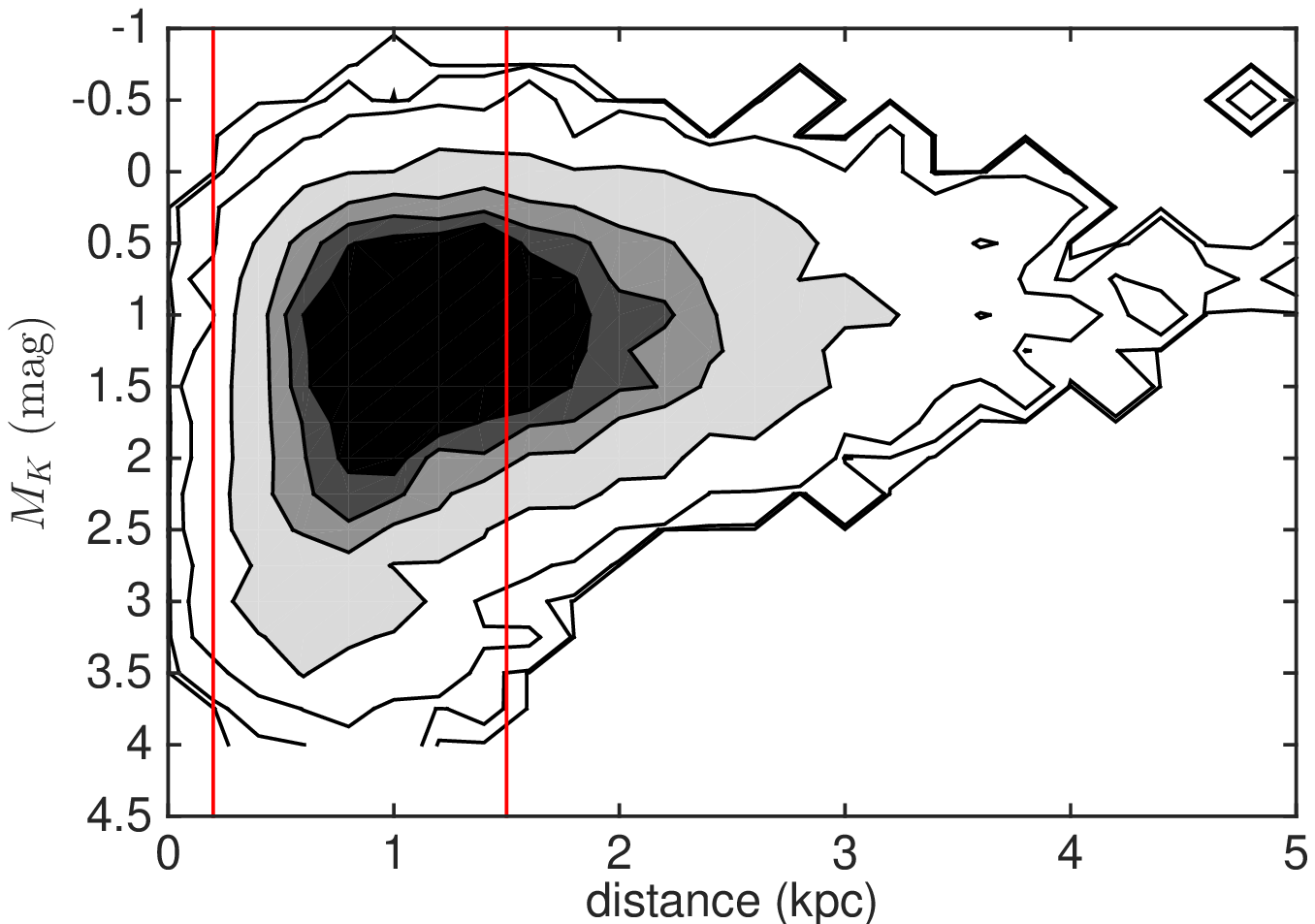}
\includegraphics[scale=0.55]{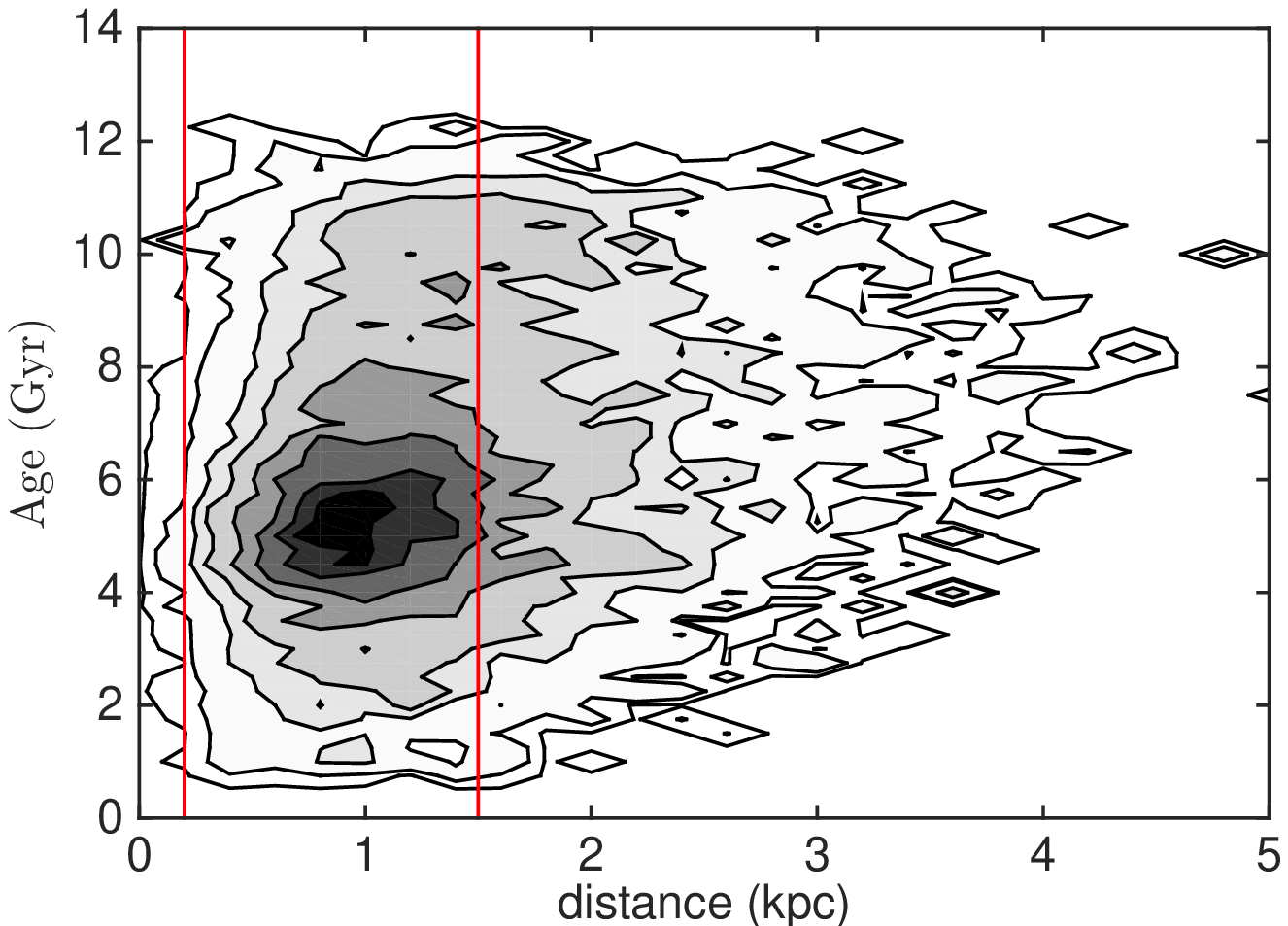}
\end{minipage}
\caption{The top panel shows the distribution of the RGB sample in \MK\ vs. distance plane. The contours indicate the stellar density with arbitrary normalization. The bottom panel shows the distribution in age vs. distance plane. The stellar sample located in between the two vertical dashed lines (0.3 and 1.5\,kpc in distance) are finally selected.}\label{fig:agehist}
\end{figure}

\begin{figure*}[ht]
\centering
\includegraphics[scale=0.5]{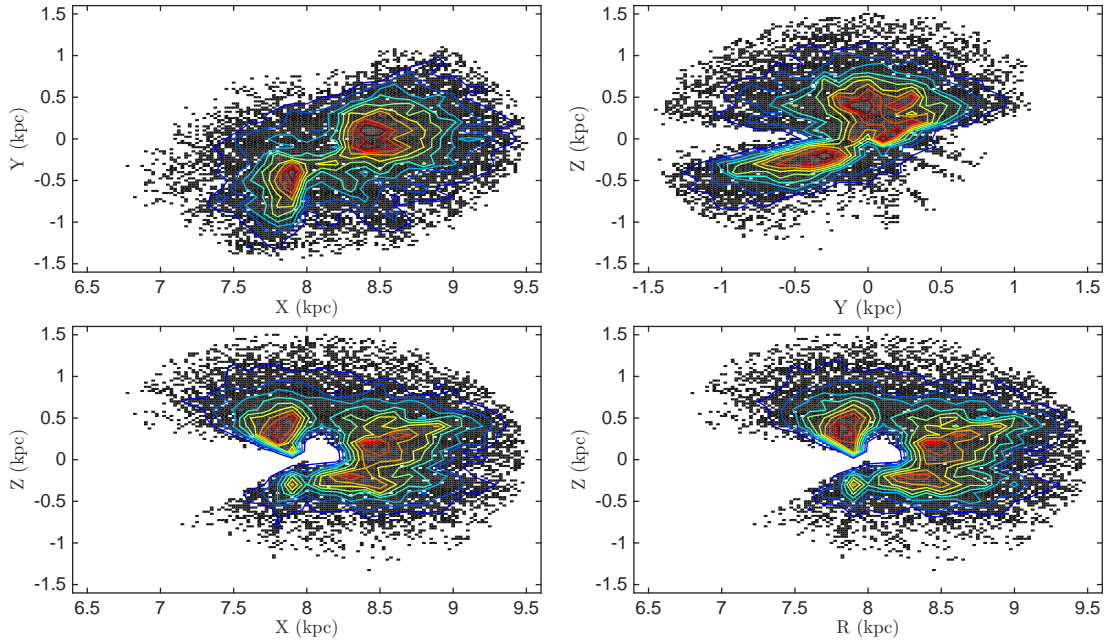}
\caption{The uncorrected spatial distribution of the final selected samples in the cylindrical coordinates, in which the Galactic center is at the origin and the negative $Y$ points to the direction of rotation.}\label{fig:xyz}
\end{figure*}
A spectroscopic survey always suffers from selection effects. Two types of selection effects have to be taken into account, one is the sampling rate of the stars and the other is the volume completeness owe to the limiting magnitude of the survey. 
Before correcting the sampling selection we have to assume that 1) the selection function of the LAMOST survey does not change the luminosity function and 2) the photometric survey data is a complete dataset~\citep{xia15b}. The former assumption is held since the targeting selection of the LAMOST survey is almost independent on the color index. For a given $i$th star, within a small solid angle and a small region in color-magnitude diagram around the star in interest, the selection correction factor is described as:
\begin{equation}\label{eq:selcorr}
w=N_{ph}{N_{sp}(M\sim M_i)\over{N_{sp,all}}},
\end{equation}
where $w$ is the selection correction factor, $N_{ph}$ the number of the photometric stars, $N_{sp}(M\sim M_i)$ the number of the spectroscopic stars with similar absolute magnitude $M$ to the $i$th star, and $N_{sp,all}$ is the the total number of the spectroscopic stars fell into the small region.
In other word, for each spectroscopic star, the selection correction factor indicates how many photometric stars with similar luminosity are located in a small surrounding volume. In principle, the smaller the selection correction factor, the fewer stars representative by the spectroscopically observed star, and verse vice. 

Specifically, we correct the selection effect in each field of view (20 square degrees) using 2MASS photometric catalog~\citep{2mass}. For each star with derived stellar parameters in the field of view, the correction factor is calculated from a small surrounding rectangle with size of $\Delta(J-K)=0.1$\,mag and $\Delta K=0.25$\,mag in $K$ vs. $J-K$ plane. The peak of the correction factors for our sample appears at around 8, meaning that most of our sample represent 8 photometric stars with similar luminosity in a small spatial volume. In order to reduce the bad correction factors due to too few data in the color--magnitude bin or bad photometry in 2MASS, we only select the stars with the correction factors between 1 and 300, which contains about 98\% of the sample.
%
In the rest of this paper, unless explicitly noted, all mean (or median) and standard deviation values for the chemodynamical properties are weighted by the selection correction factors. 

The volume incompleteness due to the limiting magnitude leads to different detecting volume for the stars with different luminosity (known as Marmquist bias). In principle, it cannot be corrected unless one has a prior knowledge of the stellar density profile and the luminosity function. The top panel of Figure~\ref{fig:agehist} shows the selection corrected distribution of the RGB stars in \MK\ vs. distance plane. The distribution of \MK\ is roughly same within 2\,kpc, while the fainter stars are progressively disappeared beyond 2\,kpc. Thus, if we take the whole sample in the study, we may substantially emphasize the bright stars in the distant volume.  Consider the volume completeness and that the stars with larger distance may have larger uncertainties in proper motions, we conservatively cut the sample at 1.5\,kpc in distance. The bottom panel shows that this cut does not induce additional selection effect in the age distribution. We finally select 17028 stars with distance between 0.3 and 1.5\,kpc. The lower limit of 0.3\,kpc is set to take into account the incompleteness for the nearest stars.
Figure~\ref{fig:xyz} shows the spatial distribution in the cylindrical coordinates for the final sample.

\section{Results}\label{sect:result}

\subsection{The resulting AMR}\label{sect:narrowstripinAMR}
\begin{figure}[htbp]
\centering
\begin{minipage}{9cm}
\centering
\includegraphics[scale=0.5]{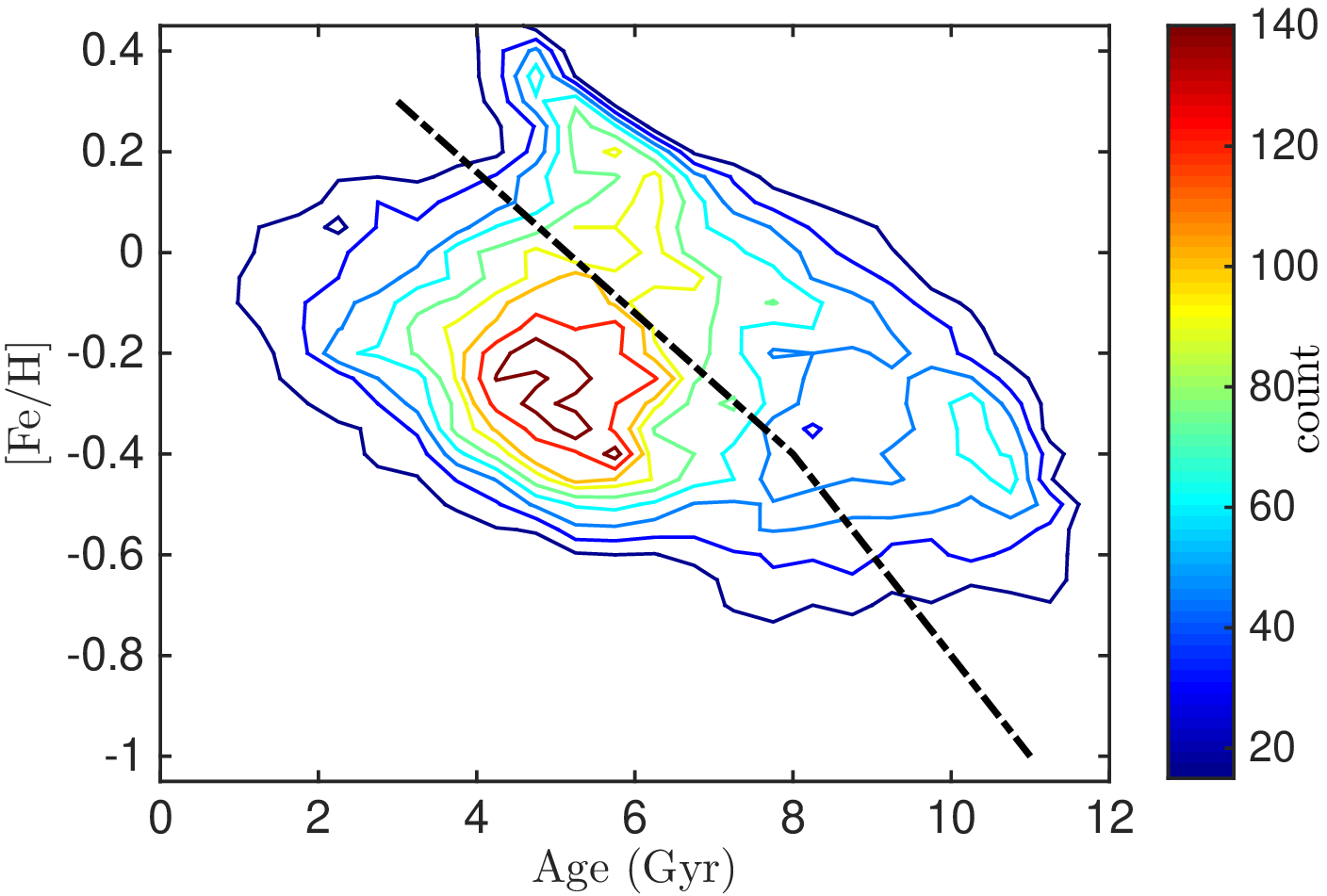}
\includegraphics[scale=0.5]{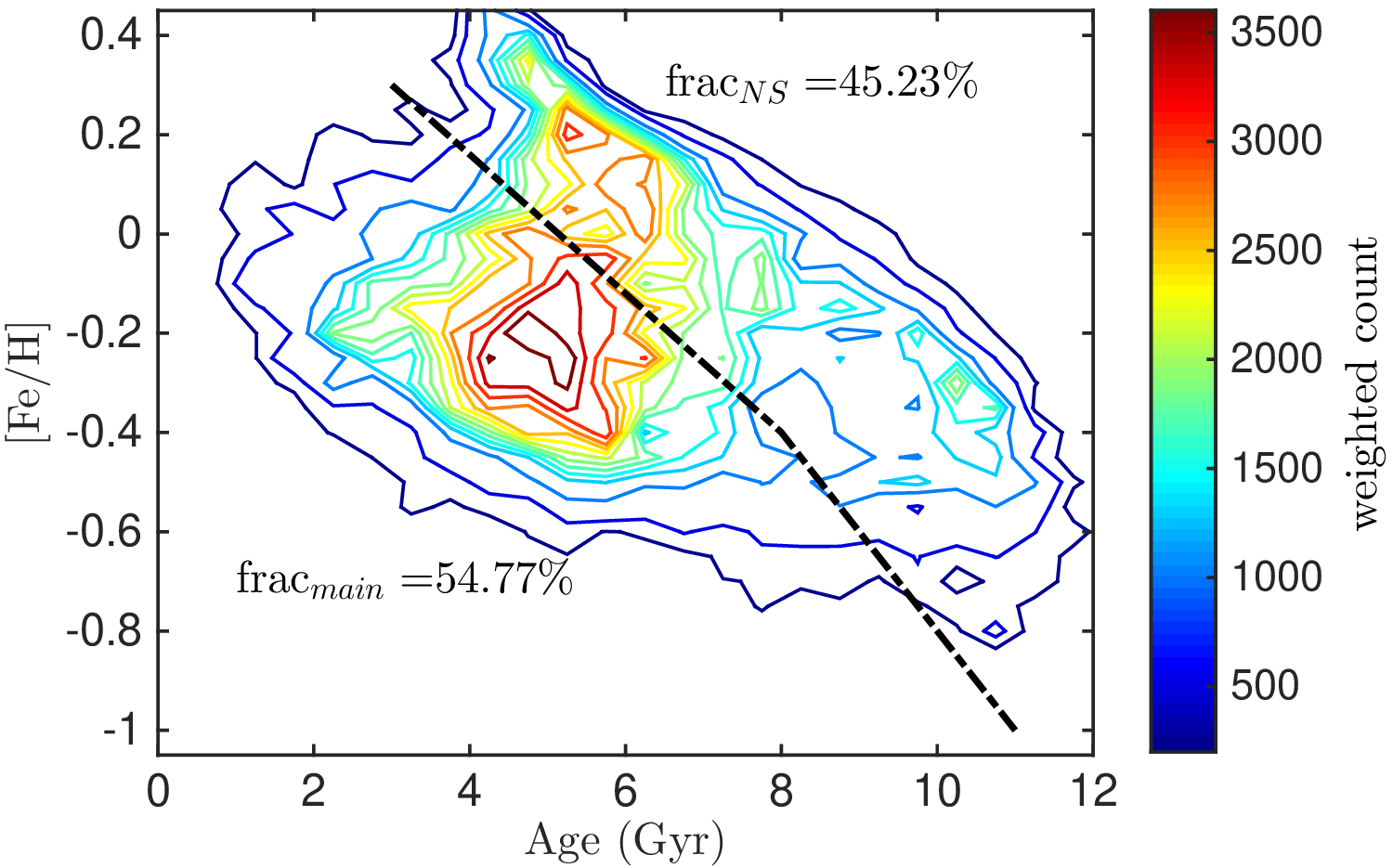}
\end{minipage}
\caption{Top panel: the selection uncorrected AM map for the RGB sample. Bottom panel: the selection corrected AM map. The bin size in both panels is $0.5$\,Gyr$\times0.05$\,dex. The dotted-dash line is the empirical separation of the \ns\ (above) and the \mn\ (below), see details in the text.}\label{fig:AMR}
\end{figure}

\begin{figure}[htbp]
\centering
\begin{minipage}{9cm}
\centering
\includegraphics[scale=0.5]{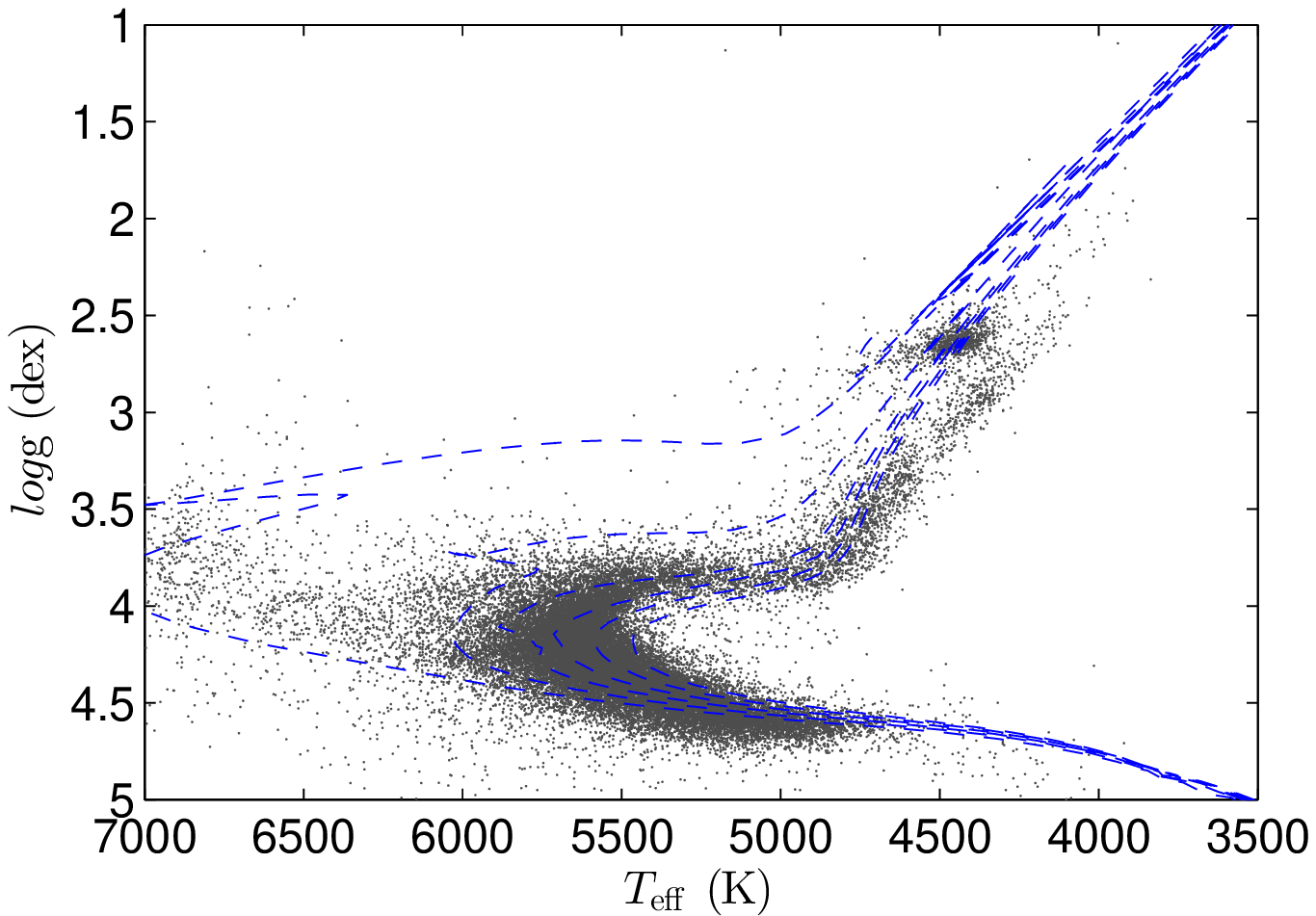}
\includegraphics[scale=0.5]{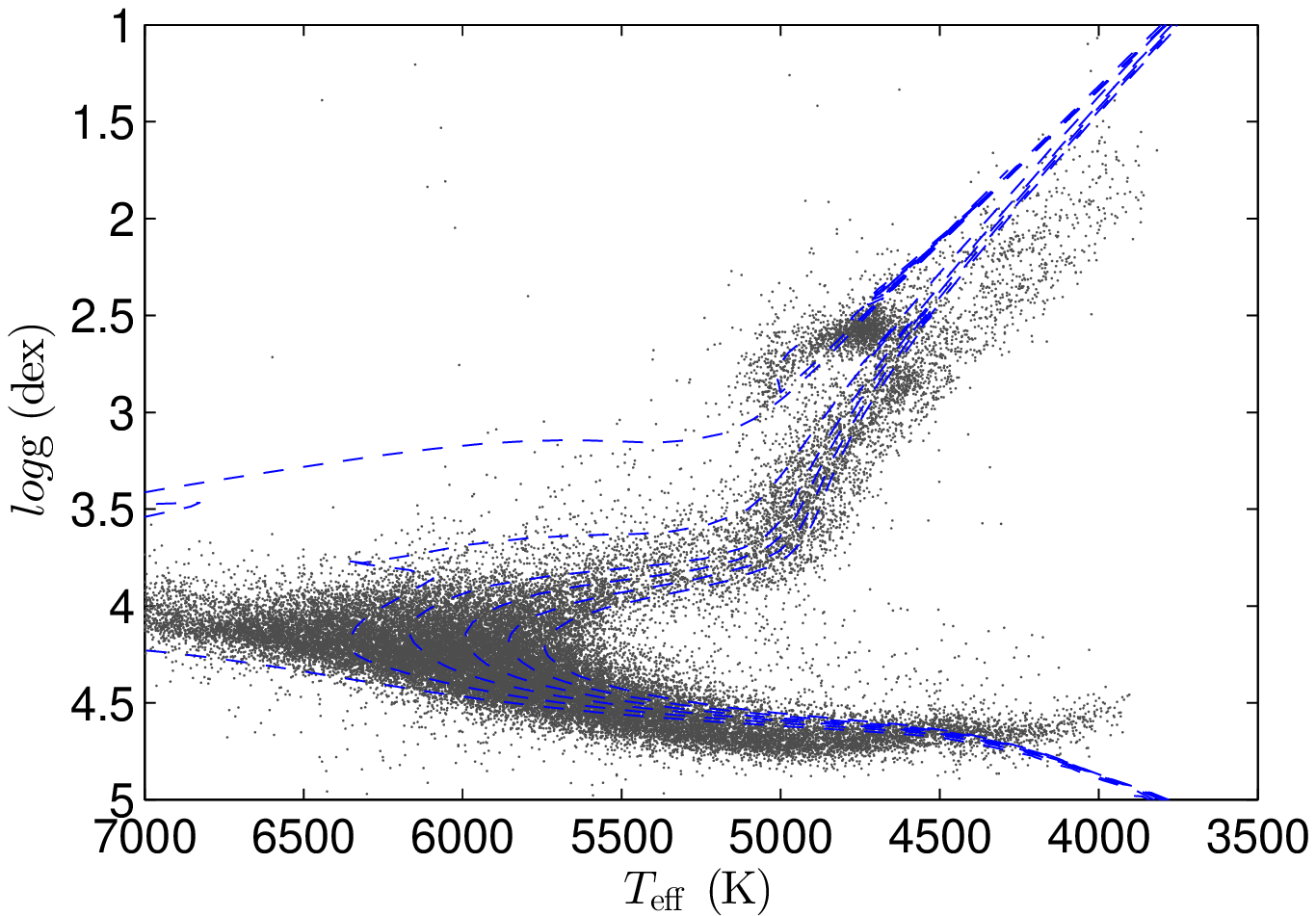}
\end{minipage}
\caption{The stars selected from the whole LAMOST DR2 catalog with \feh$>0.3$\,dex and S/N$>20$ are shown in the \logg\ vs. \teff\ plane in the top panel. The blue dashed lines are the isochrones with \feh$=0.3$\,dex and ages of 1, 3, ..., and 11\,Gyr from top to bottom, respectively. The stars selected from the LAMOST DR2 catalog with $-0.01<$\feh$<+0.01$\,dex and S/N$>20$ are shown in the bottom panel as a comparison. The superposed isochrones are at age of 1, 3, ..., and 11\,Gyr and \feh$=0.0$\,dex from top to bottom, respectively.}\label{fig:metalrichpop}
\end{figure}

Figure~\ref{fig:AMR} shows the stellar distributions in AM plane without and with selection correction (top and bottom panels, respectively). Essentially, the selection effects in the LAMOST survey seem not very severe and the selection corrected AMR is not very different with the uncorrected one. The AMR obtained in this work shows the very similar pattern to the previous works based on small sets of high spectral resolution spectra~\citep{edvardsson93,haywood13,bergemann14}. The AMR is very broad and flat with almost no stars showing up in the very old and metal-rich region as well as the very young and metal-poor region. This is likely the selection effect of the stellar populations discussed by~\citet{bergemann14}. Because our sample has been well taken into account the selection effects from targeting and the volume completeness, the selection corrected distribution of the AM map reflects the real stellar density in the volume of 0.3--1.5\,kpc around the Sun. Therefore, it can be also used to constrain the evolution of the spatial structure of the disk (Chen et al. in preparation).

Moreover, the AM map in figure~\ref{fig:AMR} shows a relatively isolated narrow stripe substructure located from (age, \feh)$\sim$(5, 0.4) to (10\,Gyr, -0.4\,dex) (the region above the dotted-dash line). The supersolar metallicity stars with \feh$>0.3$\,dex forms the head of the narrow stripe at about 4\,Gyr. A shallow gap located at about (6\,Gyr, 0.0\,dex) separates the narrow stripe from the rest stars. And it shows a prominent tail at about (10\,Gyr, -0.4\,dex). The narrow stripe is quite unusual such that : 1) the very metal-rich stars in this substructure are not young and 2) its tendency in AM map is very like an evolutionary track of a stellar population, compared to the theoretical Galactic evolution models~\citep[e.g.][]{minchev13}. 

\subsection{Verification of the \ns}\label{sect:verifynarrowstripe}
The narrow stripe could be either an artificial effect due to the systematic bias in age and/or \feh\ estimations or an interestingly real substructure never been reported in previous studies. Since almost all the supersolar metallicity stars are located in the narrow stripe, we select them as the tracer to investigate whether the substructure is artificial or real. We select all the stars, not only the RGB, but also the main-sequence and SGB, with \feh$>0.3$\,dex and S/N$>20$ from the LAMOST DR2 data. They are displayed in \logg\ vs. \teff\ plane superposed by the theoretical isochrones with \feh$=0.3$\,dex and age of 1, 3, ..., and 11\,Gyr in the top panel of Figure~\ref{fig:metalrichpop}. We also show the similar plot in the bottom panel of the figure for the stars with the solar abundance ($-0.01<$\feh$<+0.01$\,dex) and S/N$>20$ for comparison. It is quite clear that the TO/SGB stars in the supersolar metallicity subsample are mostly older than 5\,Gyr, which is qualitatively consistent with the supersolar metallicity RGB stars in figure~\ref{fig:AMR}. As a comparison, the bottom panel shows that the TO/SGB stars of the solar-abundance subsample extend broadly from 1 to 11\,Gyr\footnote{The top RGB stars of LAMOST are not consistent with the isochrones because the pipeline overestimates by $\sim0.5$\,dex in \logg\ in this region~\citep{liu15}}. This is evident that the supersolar metallicity stars in the solar neighborhood are indeed old stars. Since the most metal-rich stars is a part of the isolated narrow stripe substructure in the AM map, it consequently confirms that the narrow stripe is not an artificial but a real feature. It is noted that, recently, \citet{kordopatis15} found that the supersolar metallicity stars in the solar neighborhood is not very young population, which is well consistent with our finding. 
It is also noted that there are a few metal-rich stars located in the similar position in AM plane with slightly larger ages in the sample of~\citet{bergemann14}.

In order to reveal more properties of the narrow stripe stars, we empirically split out the RGB stars with the dotted-dash line, which basically follows the shallow gap between the narrow stripe and the rest stars, as shown in figure~\ref{fig:AMR}. The stars located above the empirical separation line are identified as the \ns. And the stars located below the line are denoted as the \mn\ stars. 

It is for certain that such a bold separation must lead to some contamination. However, as we demonstrate in next sections, carefully separating the stars with the help of the kinematic features, we can cleanly derive the chemodynamical properties for the \ns\ population. 

\subsection{Velocity distribution of the \ns}\label{sect:velstruct}

\begin{figure*}[htbp]
\centering
\includegraphics[scale=0.6]{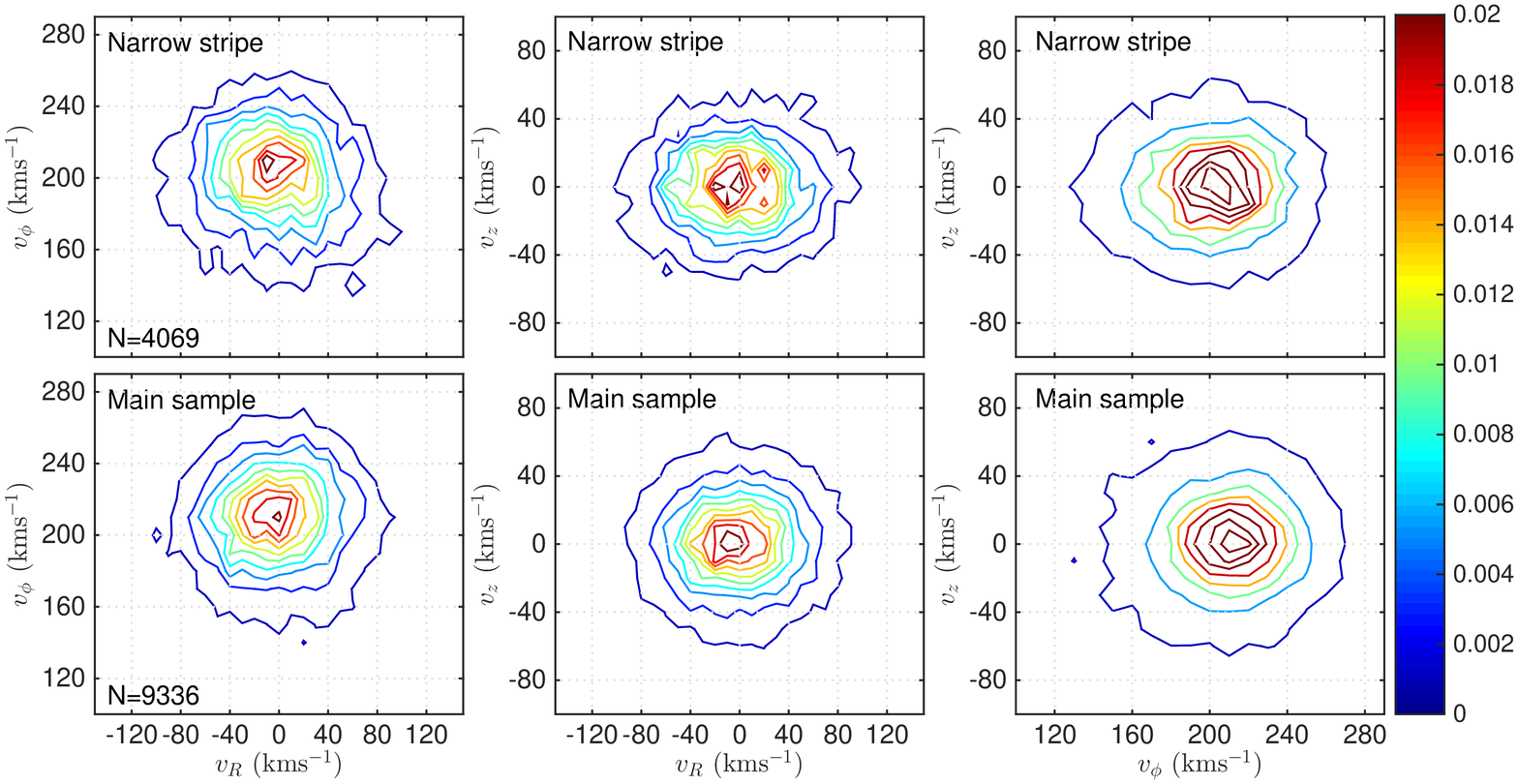}
\caption{The top panels are the projections of 3D velocity distribution in \vphi\ vs. \vrad\ (left), \vz\ vs. \vrad\ (middle), and \vz\ vs. \vphi\ (right) for the \ns\ stars. The bottom panels are the velocity distribution for the \mn. The contours are arbitrarily normalized in each distribution. 
}\label{fig:velmrich}
\end{figure*}

\begin{figure*}[htbp]
\centering
\includegraphics[scale=0.6]{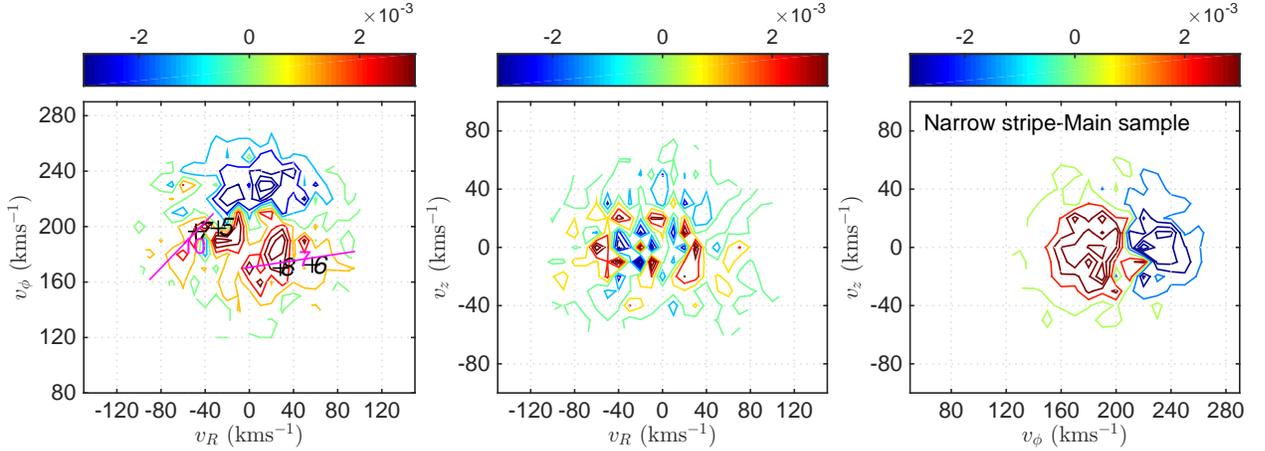}
\caption{The contours show the residual 3D velocity distribution for the \ns\ subtracted by that for the \mn. From left to right panels are the projections in \vphi\ vs. \vrad, \vz\ vs. \vrad, and \vz\ vs. \vphi, respectively. The magenta lines with numbers indicate the substructure \#7 (Hercules stream) and substructure \#10 from~\citet{xia15}. The black crosses with numbers indicate the Hercules stream (\#6 and \#8), Wolf 630 (\#5), and Dehnen98 (\#7) from~\citet{antoja12}.}\label{fig:velsubtract}
\end{figure*}

Figure~\ref{fig:velmrich} shows the 3 dimensional velocity distributions for both the \ns\ (top row) and the \mn\ (bottom row) with age smaller than 8\,Gyr. The age cut is to avoid the unfair comparison for different age ranges, since for the \mn\ only very few stars are located at age$>8$\,Gyr. Moreover, such a age cut can also avoid the large uncertainty of the age estimates for old stars, as mentioned in section~\ref{sect:ageestperform}.
It is obvious that the projected \vphi\ vs. \vrad\ distribution for the \ns\ in the top-left panel are more asymmetric with respect to \vrad$=0$ than that for the \mn. It contains more stars moving toward outer disk (\vrad$>0$) with slower \vphi. In the projected distribution of \vz\ vs. \vphi, the \ns\ contains more stars with smaller \vphi\ and thus demonstrates a longer tail at the left side.

The normalized 3D velocity distribution for the \ns\ is then subtracted by that for the \mn. The residual distribution is projected into \vphi\ vs. \vrad, \vz\ vs. \vrad, and \vz\ vs. \vphi\ and displayed in figure~\ref{fig:velsubtract}. The difference between the two velocity distributions is significantly enhanced after the subtraction. The positive excess (shown in red contours) in the residual distribution is contributed by the \ns\, while the negative excess (shown in blue contours) is contributed by the \mn. As expected, the \ns\ displays substantial excess at $160<$\vphi$<200$\,\kms, which overlaps the Hercules stream~\citep[][]{dehnen98}, Dehnen98~\citep{dehnen98,antoja12}, and the substructure \#10 (hereafter  Xia-10) in~\citet{xia15}. Moreover, another overdensity at \vphi$\sim200$\,\kms\ and \vrad$\sim-15$\,\kms, which closes to Wolf630~\citep{antoja12}, is also very prominent in the \ns. 
In the middle panel, no asymmetric structure is found in the distribution along \vz. And in the right panel, again, the \ns\ is quite prominent in smaller \vphi\ regime.

In summary, the \ns\ is not only featured in AMR, but also the dominate contributor to the velocity substructures, e.g. the Hercules stream, Dehnen98, Xia-10, and the one closed to Wolf 630, which are all in slower rotation than the \mn\ stars. 

\subsection{Identification of the velocity substructure members}
\label{sect:samplenarrowstrip}
\begin{figure*}[htbp]
\centering
\includegraphics[scale=0.6]{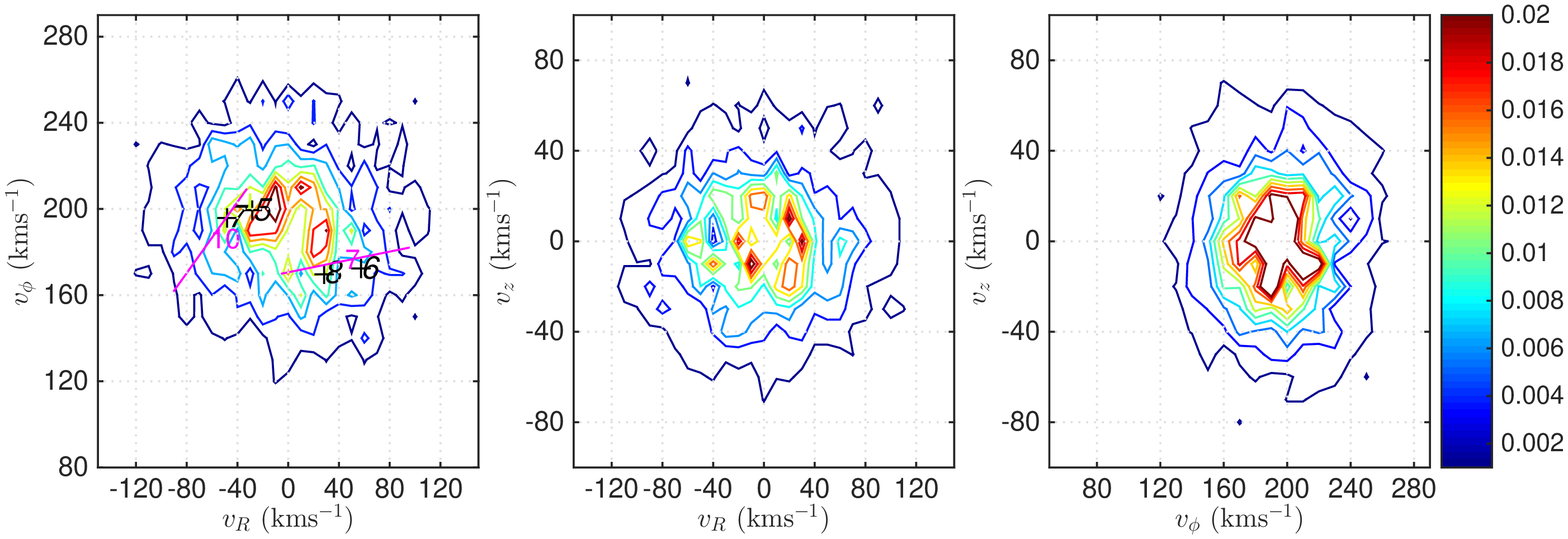}
\caption{The normalized velocity distribution for the velocity substructures after subtracting the \mn\ component from the \ns.}\label{fig:narrowstripepdf}
\end{figure*}

We identify the member stars of the velocity substructures in this section and show their chemodynamical properties in next section.
We assume that the 3D velocity distribution shown in the top panels of figure~\ref{fig:velmrich} is composed of two components, one is the smoothing distribution same as the \mn\ (the bottom row) and the other contains all the substructures. Then the velocity distribution can be decomposed as
\begin{equation}\label{eq:fracdist}
P_{NS}(\mathbf v)=P_{sub}(\mathbf v)+fP_{main}(\mathbf v),
\end{equation}
where $P_{NS}(\mathbf v)$, $P_{sub}(\mathbf v)$, and $P_{main}(\mathbf v)$ are the distributions for the \ns, the velocity substructures, and the \mn, respectively. And $f$ is the fraction of the contribution of $P_{main}(\mathbf v)$ in $P_{NS}(\mathbf v)$.
It is noted that in Figure~\ref{fig:velsubtract}, the region of \vphi$>200$\,\kms\ and $|$\vrad$|<40$\,\kms\ is dominated by the \mn. Thus, we assume that all the stars in this region in the \ns\ follow $P_{main}(\mathbf v)$. Then, $f$ can be immediately determined by comparing the number of stars within this region between the \ns\ and the \mn. Subsequently, the velocity distribution for the substructures is obtained from equation~(\ref{eq:fracdist}), as shown in figure~\ref{fig:narrowstripepdf}. It is seen that the Hercules stream, Dehnen98, Xia-10, and the one closed to Wolf 360 are all emphasized in the derived velocity distribution.

\begin{figure}[htbp]
\centering
\includegraphics[scale=0.5]{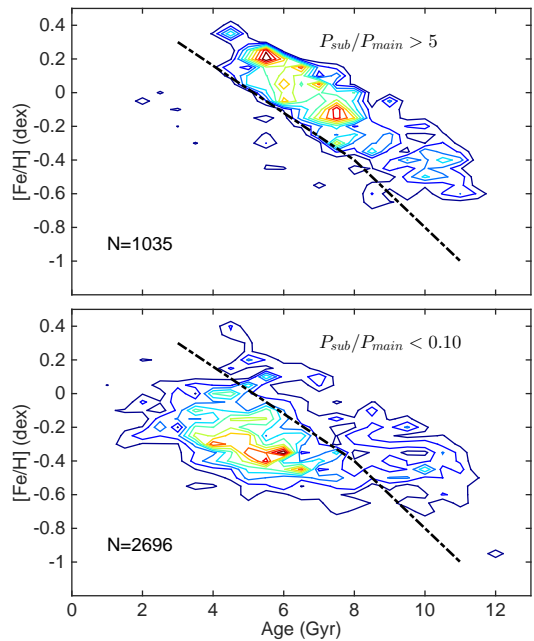}
\caption{Top panel: the AMR for the \ns\ stars with $P_{sub}/P_{main}>5$. Bottom panel: the AMR for the \mn\ stars with $P_{sub}/P_{main}<0.1$. }\label{fig:disentangleAMR}
\end{figure}

The 3D velocity distribution is equivalent with the probability density function. Thus, for a given star, we are able to calculate two probabilities: the probability to be in the substructures, $P_{sub}$, and the probability to be in the \mn, $P_{main}$. An individual star can then be identified to be in the substructures if $P_{sub}$ is significantly larger than $P_{main}$ and to be in the \mn\ if $P_{sub}$ is much smaller than $P_{main}$. Figure~\ref{fig:disentangleAMR} shows the AM map of the stars with $P_{sub}/P_{main}>5$ (top panel), which is dominated by the stars in the velocity substructures and that of the stars with $P_{sub}/P_{main}<0.1$ (bottom panel), which is dominated by the \mn\ stars. Not surprisingly, the stars in the velocity substructures are concentrated above the separation line in the \ns, while the velocity identified \mn\ stars are mostly located below. 
Therefore, the member stars of the velocity substructures with $P_{sub}/P_{main}>5$ can be the representative member stars of \ns. And the stars with $P_{sub}/P_{main}<0.1$ represent for the \mn.
Note that $P_{NS}$ and $P_{main}$ shown in figures~\ref{fig:narrowstripepdf} and~\ref{fig:velmrich} respectively are derived from the stars with age$<8$\,Gyr. Therefore, the AMR at age$>8$\,Gyr in figure~\ref{fig:disentangleAMR} is simply an extrapolation from the kinematics of younger stars and thus may not be reliable. 

\subsection{Chemodynamical properties of the \ns}
\begin{figure}[htbp]
\centering
\includegraphics[scale=0.55]{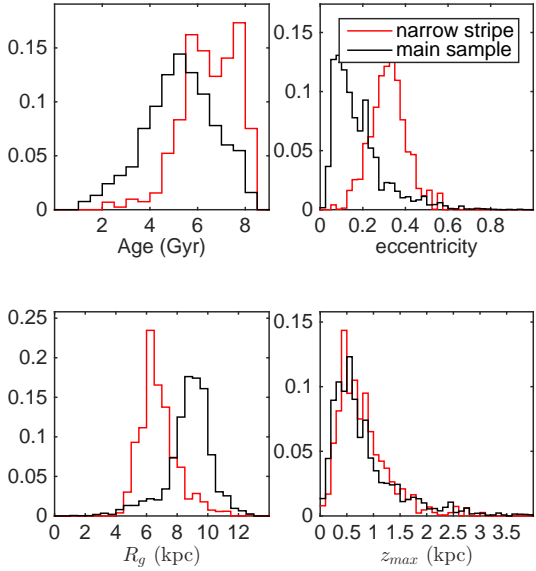}
\caption{The normalized distributions of the age (top-left), orbital eccentricity (top-right), \rg\ (bottom-left), and \zmax\ (bottom-right) for the \ns\ (red) and \mn\ (black) members identified in figure~\ref{fig:disentangleAMR}, respectively, with age$<8$\,Gyr. }\label{fig:distnarrowstripe}
\end{figure}

Figure~\ref{fig:distnarrowstripe} shows the distribution of age (top-left), eccentricity (top-right), \rg\ (bottom-left), and \zmax\ (bottom-right) for the \ns\ stars with $P_{sub}/P_{main}>5$ (red) and the \mn\ stars with $P_{sub}/P_{main}<0.1$ (black). Both groups of stars have age$<8$\,Gyr. 
In average, the \ns\ stars are older than the \mn, as shown in the top-left panel. And the age distribution of the \ns\ shows an abrupt drop at $\sim5$\,Gy with very few stars younger than this value. As a comparison, there are quite a lot of stars with ages younger than 5\,Gyr in the \mn.

In the top-right panel, the \ns\ shows quite different eccentricity distribution with the \mn. The peak of the eccentricity for the \ns\ is $\sim0.3$, which is still on near circular orbits. Moreover, the distribution of the eccentricity for the \ns\ stars seems more symmetric than that for the \mn.

In the bottom-left panel, the distribution of \rg\ for the \ns\ peaks at about 6\,kpc, smaller than the peak of the \mn\ by $\sim2.5$\,kpc. It means that although the \ns\ is located in smaller radii than the \mn, its present-day radii is still far from the central structure, e.g. bulge or bar. 

The bottom-right panel shows that, unlike the other properties, the distributions of \zmax\ for the two groups of stars are quite similar. Detailed discussions about this is in section~\ref{sect:heating}.

\section{Discussions}\label{sect:discussion}

\begin{figure}[htbp]
\centering
\includegraphics[scale=0.5]{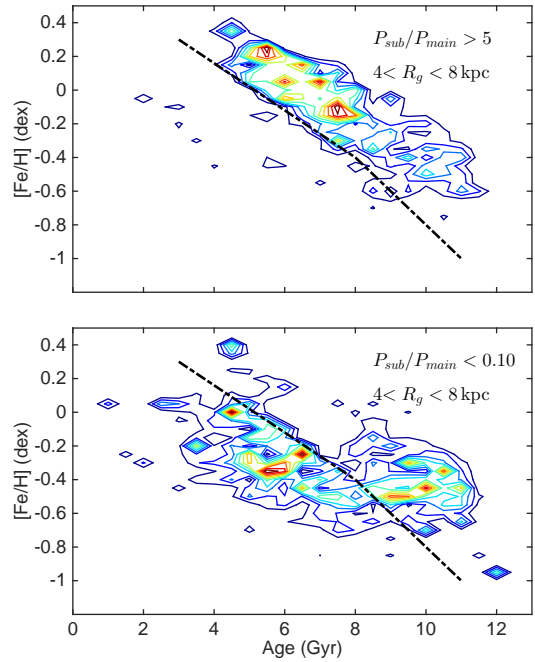}
\caption{Top panel: the contours show the AMR of the \ns\ with $P_{sub}/P_{main}>5$ and $4<$\rg$<8$\,kpc. Bottom panel: the contours show the AMR of the \mn\ stars with $P_{sub}/P_{main}<0.1$ and $4<$\rg$<8$\,kpc.}\label{fig:AMRnsmodel}
\end{figure}

\begin{figure}[htbp]
\centering
\includegraphics[scale=0.5]{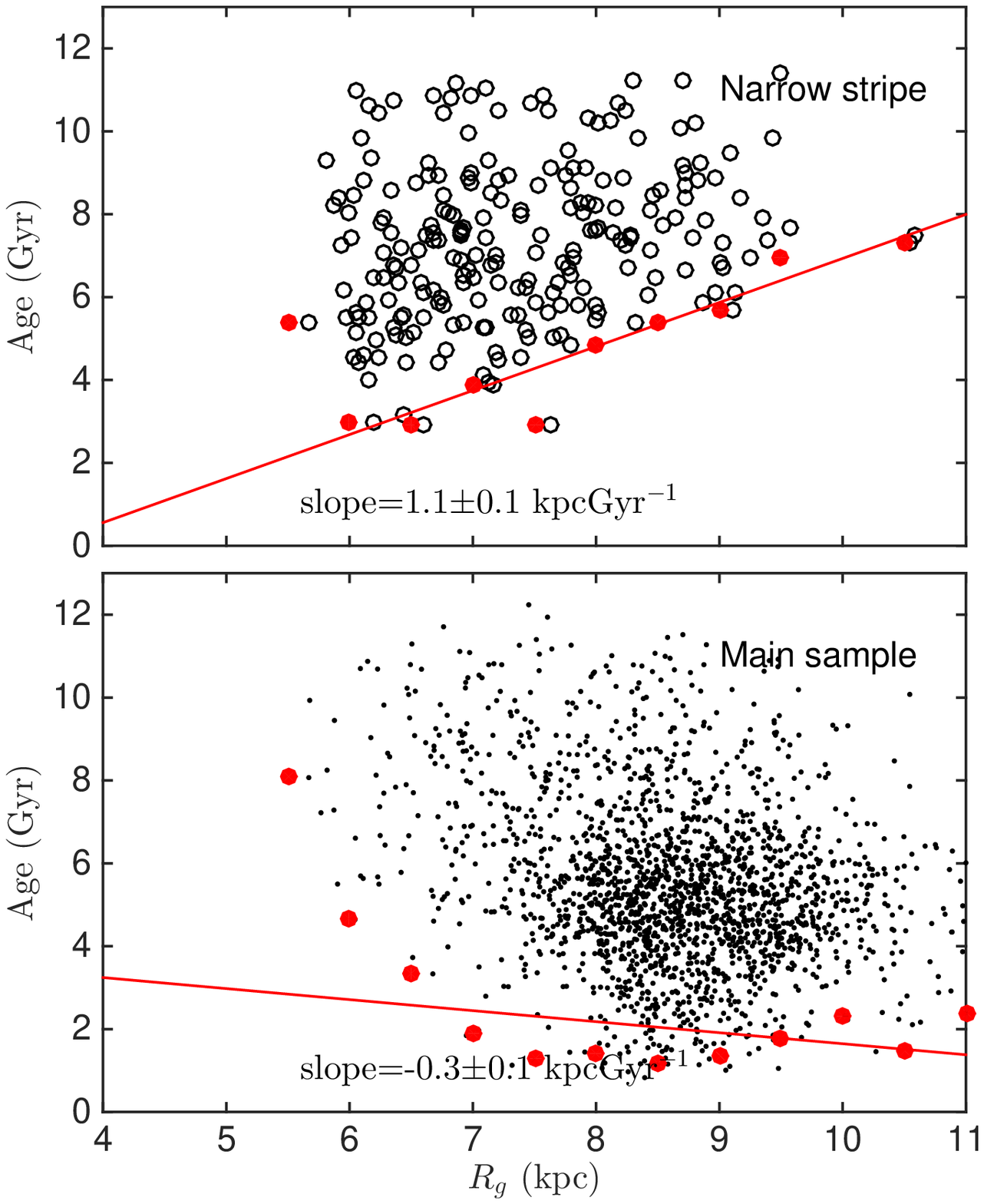}
\caption{The top panel shows the correlation between \rg\ and age for the \ns\ stars (black circles) with \ecc$<0.25$. The red filled circles show the 1st percentiles of age, which is considered as the lower limiting ages, at various \rg\ bins. They show clear correlation with \rg. The red solid line is the best linear fit of the lower limiting ages with slope of 1.1$\pm$0.1\,kpc Gyr$^{-1}$. The bottom panel shows the \mn\ stars (black dots) with \ecc$<0.25$ in age vs. \rg\ plane. The red circles and the red solid line are similar to the top panel.}\label{fig:RgAgens}
\end{figure}

\subsection{The origin of the \ns}\label{sect:narroestripeorigin}

%

From the distribution of \rg\ shown in the bottom-left panel of figure~\ref{fig:distnarrowstripe}, we find that most of the \ns\ stars are located within 4--8\,kpc. Thus, we select the \mn\ stars located in the same range of \rg\ and compare them with the \ns\ in the AM map. Figure~\ref{fig:AMRnsmodel} shows that the AMR of the \ns\ is at about 0.4-0.5\,dex above that of the \mn. This means that the \ns\ stars must be born in the much smaller radii than the \mn\ stars, although they have the same present-day \rg. According to the theoretical Galactic chemical model provided by~\citet{minchev13,minchev14}, the \ns\ is likely formed in the inner disk at around 4\,kpc. Therefore, they seem to be the radially migrated stars moving their orbits from \rg$\sim4$\,kpc to a broader range of radii centered at $\sim6$\,kpc.
Consequently, the velocity substructures, including the Hercules stream, Dehnen98, Xia-10 etc., may not be the local resonance associated with the rotating bar, but the kinematical imprints of the inner disk and later moved out due to the radial migration. We obtain that the median \vrad\ for the stars with $P_{sub}/P_{main}>5$ is $2.6^{+1.8}_{-1.9}$\,\kms. It hints that, in average, the radially migrated stars are still moving outward in present days.

Recently,~\citet{kordopatis15} found that the stars with supersolar metallicity stars are likely the radially migrated populations from about 3--5\,kpc in the inner disk. This is fully in agreement with the \ns\ sample, which contains most of the supersolar metallicity stars. However, the distribution of the eccentricity of the \ns\ stars in this work is larger than their values by about 0.1. This is probably due to the difference in sampling volume, the difference of the Galactic models used in the orbital integration, or the systematic difference in the proper motions between~\citet{kordopatis15} and this work.  

\subsection{The traveling speed of the radial migration}\label{sect:migratespeed}
If the radial migration at $\sim4$\,kpc is a continuous process, the older stars may have more time to travel to larger radii than the younger ones. Consequently, it is expected that the radially migrated stars from the same birth place (e.g. $\sim4$\,kpc) should show correlated lower limiting age with their present-day guiding-center radii. In other word, the lower limiting age would increase with \rg, since only the stars older than the lower limit are able to move to larger \rg. The top panel of figure~\ref{fig:RgAgens} shows the age vs. \rg\ correlation for the \ns\ stars with $P_{sub}/P_{main}>5$ and \ecc$<0.25.$\footnote{The cut of the eccentricity is to reduce the interference from the stars on eccentric orbits. These stars may blur the distribution of \rg\ and thus smear out the real correlation between the age and \rg\ induced by the radial migration.} The red filled circles indicate the 1st percentiles as the proxies of the lower limiting ages at various \rg\ bins. And the red solid line is the best linear fit of the 1st percentiles. It is quite clear that, indeed, the lower limiting age for the \ns\ stars does show the correlation with \rg\ and the best fit slope is $1.1\pm0.1$\,kpc Gyr$^{-1}$. As a comparison, the bottom panel of the figure shows that, for the \mn\ stars, the lower limiting age does not correlated with \rg. At \rg$<7$\,kpc, it is even anti-correlated with \rg.

The inference from the top panel of the figure is two-fold. First, the \ns\ stars are indeed the radially migrated population from the inner disk. When we extrapolate the best fit line, it approximately crosses over age$=0$\,Gyr at \rg$\sim4$\,kpc. This implies that the birth radius of the \ns\ is at about 4\,kpc, consistent with the radius inferred from the comparison of the AMR with the Galactic evolution model~\citep{minchev13,minchev14}. Second, the slope of the youngest ages, 1.1$\pm0.1$\,kpc Gyr$^{-1}$, actually stands for the mean traveling speed of the \ns\ stars, which is equivalent with $1.1\pm0.1$\,\kms. We can also obtain the traveling speed of the \ns\ stars with $P_{sub}/P_{main}>5$ from their median \vrad, $2.6^{+1.8}_{-1.9}$\,\kms. The two independent estimations of the traveling speed of the radial migration essentially agree with each other within 1-$\sigma$.

 
\subsection{The fraction of the radially migrated stars}
The efficiency of the radial migration is crucial in the sense that it reflects the importance of radial migration in the disk internal evolution. The fraction of the stars to be eventually radially migrated is an indicator of the efficiency of the radial migration. In this section, we give an approximation of the fraction for the \ns\ stars.

The stars located in the \ns\ (above the separation line) is about 45\% of the total stars in the local volume, as shown in figure~\ref{fig:AMR}. However, many stars on eccentric orbits, which may be polluted by the scattered stars, is also included. Then this number may not be the real fraction of the migrated stars. Therefore, we only consider the stars on near circular orbits in the estimation. Here we select the stars with \ecc$<0.25$. Then the fraction of the \ns\ stars to the whole sample with \ecc$<0.25$ reduces to 37\%. This value is qualitatively consistent with the  predictions from the simulations~\citep{minchev10,roskar12}. It is worthy to point out that the cut in eccentricity is more or less arbitrary. Slightly different choice around 0.25 may slightly change the fraction but may not significantly change the fact that about one third of the stars on near circular orbits are radially migrated in the solar neighborhood.

\subsection{The radial migration and disk heating}\label{sect:heating}

The effect in the disk thickening can be seen from the \zmax\ for our sample. A component may be thicker than other one if it contains more stars with larger \zmax. The distributions of \zmax\ for both the \ns\ (red) and the \mn\ (black) are compared in the bottom-right panel of figure~\ref{fig:distnarrowstripe}. Unlike the other quantities e.g. age, \rg, and \ecc, no significant difference is found in the two distributions of \zmax. It seems that the radial migration does not lead to \emph{additional} thickening of the disk given that the \mn\ is not affected by radial migration. We will focus on the disk thickening issue with the careful measurements of the spatial distributions, including the scale heights, for the mono-abundance and mono-age populations in Chen et al. (in preparation).

\section{Conclusions}
In this work, the low RGB stars, which are the good tracers of the stellar age, are selected from the LAMOST DR2 K giant sample. The age for the individual RGB stars are determined from the comparison of the stellar parameters with the theoretical isochrones. The performance assessments based on the mock datasets indicate that the uncertainty of the age estimates is about 2\,Gyr with a small fraction of stars suffering from larger systematic bias. However, we find that such systematics would not significantly distort the age--metallicity relation for the stars with age$<8$\,Gyr.

After carefully correcting the selection effects, we obtain the AMR for the sample. Surprisingly, we discover a substructure located in the narrow stripe region in the top-right part of the AM plane. We show that:
\begin{enumerate}
\item The \ns\ is composed of the supersolar metallicity stars up to \feh$\sim0.4$\,dex as well as older, lower metallicity stars.
\item The \ns\ shows strong substructures in the 3D velocity distribution, including some known moving groups, e.g., the Hercules stream, Dehnen98, Xia-10 etc.
\item With the kinematically selected sample, we are able to compare the distributions of the age, \ecc, \rg, and \zmax\ for the \ns\ with those for the \mn. They show significant differences in the distributions of age, \ecc, and \rg, but not in that of \zmax.
\item The AMR for the \ns\ favors that it is populated from the inner disk at \R$\sim4$\,kpc. However, the observed \rg\ is peaked at 6\,kpc. This large difference implies that this group of stars is likely radially migrated from 4\,kpc to the solar neighborhood. Therefore, the kinematic substructures, such as the Hercules stream, is not due to the resonance induced by the rotating bar, but probably the kinematic imprints of the inner disk around their birth place and later moved out due to radial migration. The mean eccentricity of the \ns\ stars are not like eccentric orbits, essentially in consistent with \citet{kordopatis15}. 
\item With the chemodynamical properties of the \ns, we can roughly estimate that the traveling speed of the radial migration is about 1.1$\pm0.1$\,kpc Gyr$^{-1}$, or 1.1$\pm0.1$\,\kms. This is in agreement with the median \vrad\ of $2.6^{+1.8}_{-1.9}$\,\kms\ for the \ns\ stars.
\item We also obtain that about one third stars in the solar neighborhood are radially migrated from the inner disk.
\item Finally, we do not find additional disk heating from the radially migrated stars.
\end{enumerate}

In future works, we will further investigate the chemical abundance features of the \ns\ stars and learn more details of the star formation history in the inner disk, which can better constrain the global evolution picture of the Galactic disk. The nature of the velocity substructures showing in the \ns\ also need to be carefully investigated. 

\acknowledgments
We thank Zheng Zheng, Shude Mao, and Victor Debattista for the helpful discussions and comments. This work is supported by the Strategic Priority Research Program ``The Emergence of Cosmological Structures" of the Chinese Academy of Sciences, Grant No. XDB09000000 and the National Key Basic Research Program of China 2014CB845700. CL acknowledges the National Natural Science Foundation of China (NSFC) under grants 11373032, 11333003, and U1231119. Guoshoujing Telescope (the Large Sky Area Multi-Object Fiber Spectroscopic Telescope LAMOST) is a National Major Scientific Project built by the Chinese Academy of Sciences. Funding for the project has been provided by the National Development and Reform Commission. LAMOST is operated and managed by the National Astronomical Observatories, Chinese Academy of Sciences.

\end{document}